\documentclass[preprint,nofootinbib,aps,superscriptaddress,eqsecnum]{revtex4-1}
 \pdfoutput=1
\usepackage{amsmath, amssymb, amsthm, graphicx, epsfig, fancyhdr,epsfig, slashed, mathrsfs}

\usepackage{tikzsymbols}
\usepackage{natbib}
\usepackage{xcolor}

\usepackage{tikz}
\usepackage[compat=1.1.0]{tikz-feynman}
\usepackage{cancel}
\usepackage{booktabs}
\usepackage{multirow}
\usepackage{float}
\usepackage{graphicx}
\usepackage{amsmath}
\usepackage{amsfonts}
\usepackage{amssymb}
\usepackage{float}
\usepackage{graphicx}
\usepackage{amsmath}
\usepackage{amsfonts}
\usepackage{amssymb}
\usepackage[bookmarks=true,bookmarksopen=true,
                    bookmarksnumbered=true,
                    colorlinks=true,linkcolor=black,
                    citecolor=red]{hyperref}

\usepackage{subfigure}
\usepackage{slashed}
 \usepackage{setspace} 
\usepackage{caption}
\definecolor{lime}{HTML}{A6CE39}
\DeclareRobustCommand{\orcidicon}{
	\begin{tikzpicture}
		\draw[lime, fill=lime] (0,0) 
		circle [radius=0.2] 
		node[white] {{\fontfamily{qag}\selectfont \tiny ID}};
		\draw[white, fill=white] (-0.0625,0.095) 
		circle [radius=0.007];
	\end{tikzpicture}
	\hspace{-2mm}
}

\foreach \x in {A, ..., Z}{\expandafter\xdef\csname orcid\x\endcsname{\noexpand\href{https://orcid.org/\csname orcidauthor\x\endcsname}
		{\noexpand\orcidicon}}
}

\newcommand{\be}{\begin{equation}}
	\newcommand{\ee}{\end{equation}}
\newcommand{\bea}{\begin{eqnarray}}
	\newcommand{\eea}{\end{eqnarray}}

\newcommand{\beq}{\begin{equation}}
	\newcommand{\eeq}{\end{equation}}

\usepackage[utf8]{inputenc}

\DeclareUnicodeCharacter{2212}{-}

\begin{document}

\title{ Probing Fermion-Portal Scalar Dark Matter through Charged Vector-Like Fermions at Future Muon Colliders}

\author{\small 	Songshaptak De \orcidA{}{}}
\email{songshaptak.de@ijs.si}
\affiliation{\small Jo\v{z}ef Stefan Institute, Jamova 39, 1000 Ljubljana, Slovenia \vspace{0.10cm}}

\author{\small Tapoja Jha \orcidB{}}
\email{tapoja.phy@gmail.com}
\affiliation{\small Sodankyl\"{a} Geophysical Observatory, University of Oulu, T\"{a}htel\"{a}ntie 62, Finland \vspace{0.10cm}}
\author{\small Najimuddin Khan \orcidC{}}
\email{nkhan.ph@amu.ac.in}
\affiliation{\small Department of Physics, Aligarh Muslim University, Aligarh-202002, India \vspace{0.10cm}}

\author{\small Madhurima Pandey \orcidD{}}
\email{madhurima0810@gmail.com}
\affiliation{\small Department of Physics, School of Applied Sciences and Humanities, Haldia Institute of Technology, Haldia, West Bengal-721657, India \vspace{0.10cm}}


\vspace{-1.5cm}
\begin{abstract}
We revisit a minimal fermion-portal scalar dark matter model consisting of a real singlet scalar dark matter candidate and additional vector-like singlet and doublet charged fermions stabilized by a discrete $Z_2$ symmetry. In light of the latest dark matter direct-detection constraints, the conventional Higgs-portal interaction is severely restricted, motivating a detailed investigation of fermion-mediated dark matter annihilation channels. We perform a comprehensive analysis of the model parameter space by incorporating theoretical constraints from vacuum stability and perturbative unitarity, together with experimental bounds from relic density measurements, direct-detection experiments, Higgs invisible decay searches, lepton-flavor-violating processes, and anomalous magnetic moments.
We show that the observed dark matter relic abundance can be successfully reproduced over a wide mass range through Yukawa-driven $t$- and $u$-annihilation and co-annihilation processes involving the new fermions, while remaining consistent with current direct-detection limits. Motivated by the viable parameter space, we investigate the discovery prospects of the lightest charged vector-like fermion at future muon colliders operating at center-of-mass energies of 3 TeV and 10 TeV. Focusing on the process $\mu^+\mu^- \to E_1^+E_1^- \to e^+e^- + \cancel{E}_T$, we perform a detector-level analysis including realistic Standard Model backgrounds. We demonstrate that the clean experimental environment of a muon collider provides excellent sensitivity to charged fermion masses extending into the multi-TeV regime, significantly improving the exploration prospects of this class of fermion-portal dark matter scenarios.

\end{abstract}
\keywords{Dark matter, neutrino mass and mixing, lepton flavour violation}
	\maketitle
\newpage 

\section{Introduction}

The last few decades have seen a revolution in cosmology and astrophysics. The satellite-based experiments, Wilkinson Microwave Anisotropy Probe (WMAP)~\cite{Komatsu:2014ioa, WMAP:2012fli} and Planck~\cite{Planck:2013pxb, Planck:2018jri}, using the anisotropies in Cosmic Microwave Background Radiation (CMBR), suggest that the total mass-energy of the Universe contains $69\%$ dark energy, $27\%$ dark matter, and $4\%$ ordinary matter, i.e., the Universe has more dark matter than matter. 
So far, astrophysical and cosmological data can only tell us how much dark matter is in the Universe, and that it does not interact electromagnetically or strongly. However, there is still no consensus on what it is composed of. The possibilities include mostly non-baryonic matter and partly dense baryonic matter that does not emit radiation. The non-baryonic dark matter can be grouped into three categories based on their free-streaming length, namely hot dark matter (HDM), warm dark matter (WDM), and cold dark matter (CDM).
Currently, all viable models of structure formation indicate that the CDM may dominate the Universe. It is yet very difficult to determine the constituents of CDM. Possible candidates include axions and feebly/weakly/strongly interacting massive particle(s) (FIMPs/WIMPs/SIMPs)~\cite{Hall:2009bx}.
There are various direct~\cite{LZ:2024zvo,PandaX-II:2017hlx,XENON:2018voc,PICO:2019vsc} and indirect~\cite{Daylan:2014rsa, MAGIC:2016xys,Gruber:1999yr} detection experiments are operating nowadays.
We, however, find astrophysical evidence of dark matter via a particle-radiation excess, but we still have not obtained any direct signature of dark matter.
The Large Hadron Collider (LHC)~\cite{CMS:2022dwd, ATLAS:2022vkf, DeCosa:2015bmq}, along with other collider experiments, such as Large Electron–Positron Collider (LEP)~\cite{Lundstrom:2008ai}, scrupulously search for dark matter via the missing-energy signature of the dark matter. After several years of analysis, these collider experiments can probe a window for new physics beyond the Standard Model (SM). 

Various extensions have been studied in the literature to explain the dark matter relic abundance. Extensions of the SM with additional scalar fields, including singlet, doublet, and triplet representations, have been extensively studied in the literature (see Refs.~\cite{Silveira:1985rk, Khan:2014kba, Deshpande:1977rw, Chaudhuri:2015pna, Chen:2008jg, Khan:2016sxm}). Models incorporating additional fermionic dark matter candidates have also received significant attention~\cite{Chaudhuri:2015pna}. Furthermore, mixed scenarios involving both extended scalar sectors and fermionic dark matter have been investigated in various works, providing rich phenomenological implications for dark matter, collider physics, neutrino, and cosmology~\cite{Dey:2025pcs, Das:2020hpd, Das:2021zea, Das:2021qqr}.

In this work, we revisit the real singlet scalar dark matter model with additional new singlet and doublet vector-like fermions~\cite{Das:2020hpd, Das:2021zea, Khan:2022kis}. We investigate the constraints on the dark matter parameter space and observe significant changes to the allowed regions once the latest direct-detection bounds are imposed~\cite{LZ:2024zvo}. We reduce the Higgs portal coupling to avoid the direct detection limits. We then present the viable parameter space in the presence of the new Yukawa couplings through additional $t$- and $u$-channel processes which involve the new fermions. In our previous work~\cite{Das:2020hpd, Das:2021zea}, we showed observed/projected reach for the direct new fermion searches at the High-Luminosity LHC (HL-LHC) ($\sqrt{s}=14~{\rm TeV}, \mathcal{L}=~3~{\rm ab^{-1}}$), which has huge Standard Model backgrounds. We have shown that the reach of the new singlet-doublet fermion mass, depending on mixing, is up to $1.0-1.5$ TeV. While these results demonstrate the capability of the HL-LHC to probe a significant portion of the parameter space, heavier states and more weakly coupled regions remain challenging to access. The proposed muon collider offers a promising opportunity to extend the exploration of such scenarios. Recent design studies propose its operation at center-of-mass energies of $\sqrt{s}=3$ and $10~\mathrm{TeV}$ with integrated luminosities of $1~\mathrm{ab}^{-1}$ and $10~\mathrm{ab}^{-1}$, respectively~\cite{InternationalMuonCollider:2025sys}. In contrast to hadron colliders, a muon collider provides a much cleaner experimental environment with a well-defined initial state and substantially reduced QCD backgrounds. Furthermore, it allows the full centre-of-mass energy to be exploited in hard scattering processes. These features significantly enhance its sensitivity to electroweakly interacting new particles and make the muon collider an ideal facility for probing dark matter motivated extensions of the SM. Taking inspirations from these developments, we set our goal to search for a projected region of our model parameters in the proposed muon collider. The main goal of this study is to determine the regions of parametric spaces that can be probed via the direct production of the new fermionic states at a high-energy muon collider. We then assess the extent at which the muon collider can surpass the mass reach of the HL-LHC. Finally, we examine whether the parameter regions accessible at the muon collider ar consistent with the existing dark matter constraints.

The work is organized as follows. We provide the complete model description along with possible theoretical and experimental constraints in section~\ref{sec2}. We present a detailed numerical analysis of dark matter from relic density and direct detection in section~\ref{sec:dm1}. We explore the collider study in section~\ref{sec:col}. Finally, we conclude our results in section ~\ref{sec:conc}.
\section{Model}\label{sec2}
\noindent
We consider an extension of the Standard Model that consists of a vector-like charged fermion doublet $F_{D}$, a vector-like fermion singlet $E_{S}^{-}$ with electric charge $-$1, and a real singlet scalar $S$ along with the SM particles. The $F_{D}$ doublet is given by $(X_{1}^{0}, E_{D}^{-1})^{T}$, where $X_1^0$ denotes the electrically neutral component of the doublet, while $E_D^{-}$ represents the charged component with electric charge -1. Since the new fermions are vector-like, they do not make any additional contribution to the gauge anomalies~\cite{pal2014introductory, Pisano:1993es}. We impose a discrete $Z_{2}$ symmetry in the theory, where these new particles are assumed to be $Z_{2}$-odd. The SM are assumed to be $Z_{2}$ even and hence there is no mixing in the mass matrices between the SM and the new particles of the theory. The lightest neutral new particle acts as the viable dark matter (DM) candidate. The corresponding Lagrangian is given as~\cite{Das:2020hpd,Das:2021zea}
\begin{equation}
	\mathcal{L}_{\rm tot}=\mathcal{L_{\rm SM}}+\mathcal{L_S}+\mathcal{L}_{\rm f}+\mathcal{L}_{\rm int}.
\end{equation}
The expressions of different parts of the Lagrangian are given by,
\begin{eqnarray}
	\mathcal{L_S}&=&\frac{1}{2}|\partial_{\mu} s|^2-\frac{1}{2}\kappa s^2\phi^2-\frac{1}{4} m_{s}^2 s^2-\frac{\lambda_s}{4!} s^4\label{eq:scalar},\\
     \mathcal{L}_{\rm f}&=&\overline{F}_D\gamma^{\mu}D_{\mu}F_D+\overline{E}_S\gamma^{\mu}D_{\mu} E_S-M_{ND}\overline{F}_D F_D - M_{NS}\overline{E}_S E_S,\\
	\mathcal{L}_{\rm int}&=&-Y_{N}\overline{F}_D\phi E_S - Y_{fi} \overline{\psi}_{i,L} F_D \, s  - \, Y_{fi}^{\prime }\, \overline{l}_{i,R} E_S \, s + {\rm h.c.}.~\label{lint}
\end{eqnarray} 
In the above equations, $D_{\mu}$ corresponds to the covariant derivative of the singlet and doublet fermions, ${\psi}_{i,L}$ and $l_{i,R}$ represent the conventional SM left-handed doublet and the right-handed singlet charged leptons, where the indices $i=e,\mu,\tau$ stand for three generations. $\kappa$ stands for Higgs portal coupling, $M_{ND}$ and $M_{NS}$ are the masses of the new fermionic doublet and singlet, respectively, and $Y$s are the new Yukawa matrices. Here, $\phi$ is the standard Higgs doublet given by $( \phi^+, \frac{v+h+i\chi^0}{\sqrt{2}})^T$. The corresponding potential can be written as,  $V_{\rm SM}(\phi)=-\mu^2\phi^2+\lambda\phi^4$. The vacuum expectation value is given by $v=246.221$ GeV. Due to symmetry breaking, the mixing between charged doublet, $E_D^\pm$ and singlet fermionic states, $E_S^\pm$ at tree level arises. The corresponding mixing matrix is given by
\begin{eqnarray}
	\mathcal{M}=\begin{pmatrix}
		M_{ND}& M_X\\M_X^{\dagger}&M_{NS}\\
	\end{pmatrix},
	\label{eq:mass1}
\end{eqnarray}
where, the off-diagonal term $M_{X}$ is proportional to $v$ and given by $v Y_{N}/\sqrt{2}.$ We obtain the physical eigenstates with corresponding eigenvalues by diagonalizing the above mass matrix by the following rotation in the ($E_D^\pm$,  $E_S^\pm$) plane. The diagonalization has been achieved in the following way,

\begin{eqnarray}
	\begin{pmatrix}
		E_1^\pm\\E_2^\pm\\
	\end{pmatrix}=\begin{pmatrix}
		\cos\beta&\sin\beta\\-\sin\beta&\cos\beta\\
	\end{pmatrix}\begin{pmatrix}
		E_D^\pm\\E_S^\pm\\
	\end{pmatrix}, {~\rm with~} \tan 2 \beta = \frac{2 M_X}{M_{NS}-M_{ND}}.
\end{eqnarray}
Taking into account the relation ($M_{NS}-M_{ND}) \gg M_X$, we obtain the eigenvalues as,
\begin{eqnarray}
	M_{E_1^\pm} = M_{ND} - \frac{2 (M_X)^2}{M_{NS}-M_{ND}},\,
	M_{E_2^\pm} = M_{NS} + \frac{2 (M_X)^2}{M_{NS}-M_{ND}}.\nonumber
\end{eqnarray}
The masses of  the scalar fields and the neutral fermion are given by the following expressions,
\begin{eqnarray}
	M_{s}^2=\frac{m_{s}^2+\kappa v^2}{2}~, \  M_h^2= 2\lambda v^2,\, \,\text{and} \,\, M_{X_1^0}=M_{ND}. 
	\label{eq:mass}
\end{eqnarray}
From the eigenvalue equations we can see $M_{E_1^\pm}<M_{X_1^0}<M_{E_2^\pm}$. Evidently, the neutral fermion cannot serve as a viable DM candidate. Setting the condition $M_{s} < M_{E_1^\pm}$, we can have the singlet neutral scalar $s$ as the viable DM candidate of the model. We will discuss the phenomenology of  DM in detail in section~\ref{sec:dm1}.

\noindent
The parameter space of the model is constrained from different theoretical considerations, e.g., vacuum stability, metastability, unitarity of the scattering processes and also from DM relic density~\cite{Planck:2018vyg} and direct detection cross section~\cite{LZ:2024zvo}. The measurements of Higgs invisible decay width and signal strength at LHC set further constraints~\cite{CMS:2022ley, ATLAS:2023dnm}.

\noindent
In the next section,  we discuss the constraints obtained from stability, unitarity, lepton flavour violation ($\mu\rightarrow e\gamma$) along with the anomalous magnetic moment.

\subsection{Absolute stability and unitary constraints and Inflation}\label{sec:lfv}
\noindent
To obtain the stability condition, the potential should be bounded from below. In our scenario, for very large value of scalar fields $h, s \gg v$, the scalar potential (eqn.~(\ref{eq:scalar})) can be given as,
\begin{eqnarray}
V(h,~s) = \frac{1}{4}\left( \sqrt{\lambda} h^2 - \sqrt{\frac{\lambda_s}{6}} s^2 \right)^2 + \frac{1}{4}\left(\kappa +   \sqrt{\frac{2 \lambda \lambda_s}{3}}\right) h^2 s^2.
\label{scalpotstability}
\end{eqnarray}

\noindent
Therefore, the absolute stability conditions are obtained as,
\begin{equation}
\lambda(\Lambda) > 0, \quad \lambda_s(\Lambda) > 0 \quad {\rm and} \quad \kappa(\Lambda) + \sqrt{\frac{2 \lambda(\Lambda) \lambda_s(\Lambda)}{3}} > 0.
\end{equation}

\noindent
Using Renormalization Group (RG) equations~\cite{Khan:2012zw}, the coupling constants are evaluated at a scale $\Lambda$. It should be taken into account that RG equations can modify the scalar quartic couplings, so that $\lambda$ and/or $\lambda_s$ can become negative at some energy scale. We therefore need to carefully account for the constraints arising from the stability condition, following Ref.~\cite{Khan:2012zw}. Furthermore, to ensure that the model remains perturbatively calculable, the absolute values of the quartic couplings must satisfy the condition $|\lambda_i| \leq 4\pi$.

\noindent
To remain predictive at high energy scale, the constraints from the unitarity of the scattering matrix ($S$-matrix)~\cite{Lee:1977eg, Cynolter:2004cq} should be maintained. The scattering cross-section for any process given by the Born approximation can be written as $\sigma = (16 \pi)/s \sum_{l=1}^\infty (2l+1) |a_l(s)|^2$. Here, $s$ is the Mandelstam variable, $a_l(s)$ are the coefficients of partial waves with specific angular momenta $l$.
At very high energies, the contributions from the processes involving the s-, t-, and u-channel diagrams become negligible, and the S-matrix is determined solely by processes involving quartic interactions. According to the equivalence theorem, the unitarity condition  ${\rm Re}[a_l(s)]< 1/2$ can then be translated into the requirement that the eigenvalues of the $S$ matrix must not exceed $8\pi$. The Unitarity condition of the following scattering processes  $\chi^0 \chi^0 \rightarrow \chi^0 \chi^0$, $\phi^+\phi^- \rightarrow \phi^+\phi^-$ and $(\chi^{0},\phi^{\pm}) s \rightarrow (\chi^{0},\phi^{\pm}) s$ can be mentioned as 
\begin{equation}
\lambda (\Lambda)\, 
 <\, \frac{8\pi}{3} \, ; \,\,\, |\kappa(\Lambda)|\, < \,8\pi.
 \label{eq:pert}
 \end{equation}
\noindent
The most stringent constraints from unitarity come from the scattering processes $hh \rightarrow hh $,  $hh \rightarrow ss $ and  $ss \rightarrow ss $~\cite{Cynolter:2004cq} which is given by
\begin{equation}
\Big| 12 {\lambda(\Lambda)}+{\lambda_s(\Lambda)} \pm \sqrt{16 \kappa(\Lambda)^2+({\lambda_s(\Lambda)}-12 {\lambda(\Lambda))^2}}\Big| \leq 32 \pi.
\label{eq:unitary}
\end{equation}

In the present model~\cite{Das:2020hpd,Das:2021zea,Khan:2022kis}, the additional scalar sector allows both $\lambda$ and $\lambda_s$ to remain positive up to high scales. Consequently, the Higgs and/or the singlet scalar can act as viable inflaton candidates~\cite{Lerner:2009xg, Lebedev:2011aq}. 
Using the RG-evolved scalar quartic couplings at the GUT scale together with non-minimal couplings $\zeta_{h,S}$ to the Ricci scalar, we obtain successful inflation consistent with the observed number of $e$-folds and the current bounds on inflationary observables such as scalar spectral index $n_s$, tensor-to-scalar ratio $r$, and running of the scalar spectral index $n_{rs}$ from Planck data~\cite{Planck:2018jri}.

\subsection{Lepton flavour violation ($\mu\rightarrow e\gamma$) and anomalous magnetic moment}
\noindent
To explain the anomalous magnetic moment of muon, the model requires the Yukawa couplings to be very high, greater than $\sqrt{4\pi}$~\cite{Abi:2021gix}; e.g., high BSM masses ($\mathcal{O}( 1 - 3)$ TeV) and high value of Yukawa coupling can given anomalous magnetic moment as $\delta a_\mu \sim 25\times10^{-11}$. These combinations of couplings certainly violate perturbativity and also the constraint coming from lepton flavour violation (LFV), which is BR$(\mu \rightarrow e\gamma) < 4.2\times10^{-13}$ at $90\%$ C.L.~\cite{Baldini:2018nnn}. By opting the second generation Yukawa couplings extremely small $\sim 10^{-3}$ and other couplings of the order of $\mathcal{O}(1)$, we can evade both constraints (LFV and anomalous magnetic moment)~\cite{Das:2020hpd, Das:2021zea} in this model.

\section{Dark matter}
\label{sec:dm1}
 As shown in Refs.~\cite{Das:2020hpd,Das:2021zea,Khan:2022kis}, the lightest $Z_2$-odd singlet scalar $s$ provide a viable dark matter (DM) candidate. Depending on the choice of parameters, the observed relic abundance can be generated via either Freeze-out or Freeze-in. In the present analysis, we focus on the Freeze-out regime and re-examine the allowed parameter space by incorporating the latest bounds from dark matter direct-detection experiments together with other phenomenological constraints. We also explore the implications of the model for future muon collider studies.

\begin{figure}[h!]
	\begin{center}
		\subfigure[]{
			\includegraphics[scale=0.4]{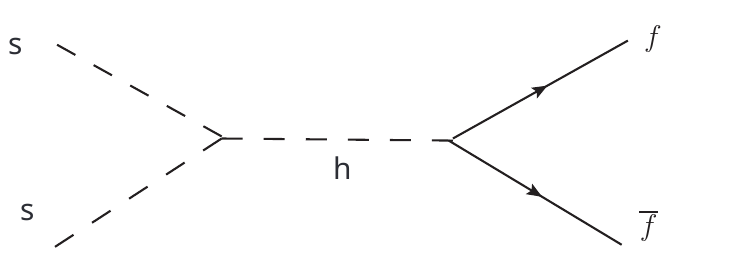}}
		\hskip 1pt
		\subfigure[]{
			\includegraphics[scale=0.4]{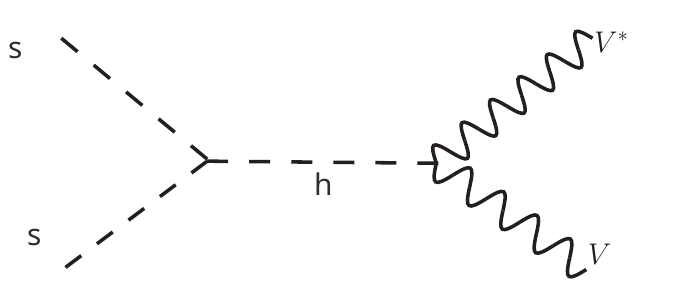}}
		\subfigure[]{
			\includegraphics[scale=0.32]{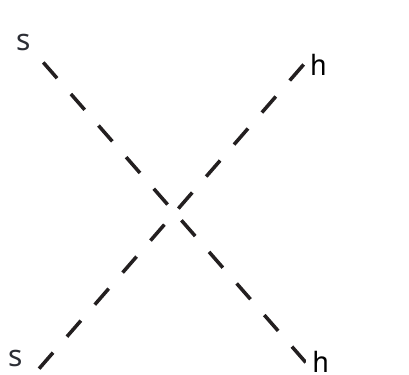}}
		\hskip 1pt
		\subfigure[]{
			\includegraphics[scale=0.4]{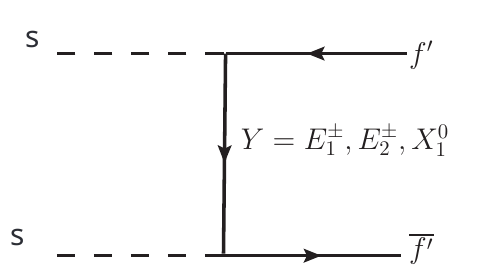}}
		\caption{ \it The dark matter annihilation diagrams give the relic density. $V$ stands for gauge bosons $W,Z$. Here, $f^\prime$ denotes the SM leptons, while $f$ represents the SM fermions, including both leptons and quarks}
		\label{fig:DarkAn}
	\end{center}
\end{figure}
\begin{figure}[h!]
	\begin{center}
		\subfigure[]{
			\includegraphics[scale=0.6]{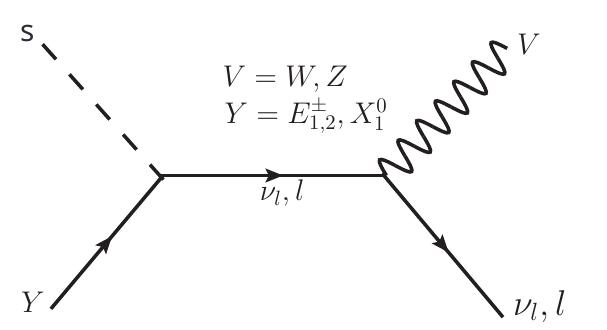}}
		\hskip 1pt
		\subfigure[]{
			\includegraphics[scale=0.6]{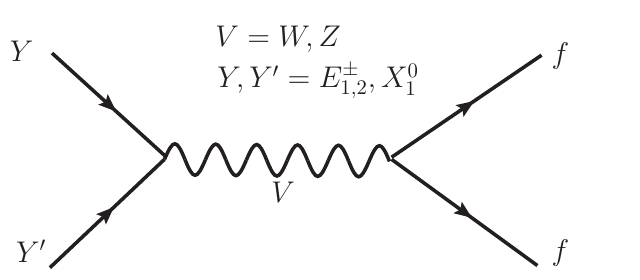}}
		\subfigure[]{
			\includegraphics[scale=0.6]{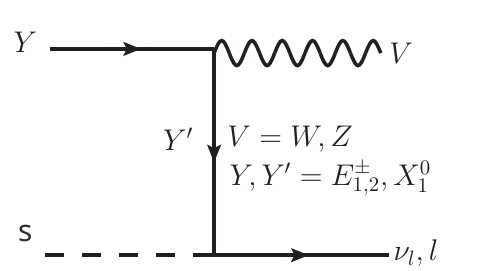}}
		\caption{ \it The co-annihilation and annihilation diagrams of the dark matter and the other $Z_2$-odd fermion fields. $f$ are standard model leptons and quarks.
		}
		\label{fig:DarkCoan}
	\end{center}
\end{figure}

In this scenario, the DM relic density is determined by annihilation and co-annihilation processes involving the scalar $s$ and the additional $Z_2$-odd fermions. The dominant contributions arise from the Higgs portal interaction $\kappa$, which mediates the $s$-channel and contact interaction processes shown in Figs.~\ref{fig:DarkAn}(a)--(c). Additional contributions originate from the Yukawa interaction $Y_{fi}$ through the $t$-channel process displayed in Fig.~\ref{fig:DarkAn}(d). Co-annihilation channels involving the heavier fermionic states, illustrated in Fig.~\ref{fig:DarkCoan}, can also become important in certain regions of parameter space. To study DM phenomenology, the model is implemented in {\tt FeynRules}~\cite{Alloul:2013bka} and interfaced with {\tt micrOMEGAs}~\cite{Belanger:2018mqt} for the relic-density calculations. The numerical results are independently verified using the {\tt SARAH-4.14.3}~\cite{Staub:2013tta, Staub:2015kfa} framework together with the spectrum generator {\tt SPheno-4.0.3}~\cite{Porod:2011nf}, which is further connected to {\tt micrOMEGAs}.

The Higgs portal interaction contributes through the couplings $g_{hss}=\kappa v,~g_{hhss}=\kappa$,
and is strongly constrained by present direct-detection searches. In contrast, the Yukawa interactions $Y_{fi}$ and $Y_{fi}^{\prime}$ mainly affect the relic density through $t,u$-channel annihilation and co-annihilation processes. Their contributions are governed by the effective couplings
$g_{siE_1^\pm}~ (=A_{fi}=|\cos\beta\,Y_{fi}+\sin\beta\,Y_{fi}^{\prime}|$),
$g_{siE_2^\pm}~(=B_{fi}=|\sin\beta\,Y_{fi}+\cos\beta\,Y_{fi}^{\prime}|$),
and $g_{s\nu_i X_1^0}~(=C_{fi}=|Y_{fi}|)$, where $i=e,\mu,\tau$.

\begin{figure}[h!]
	\centering
    	\includegraphics[scale=.3]{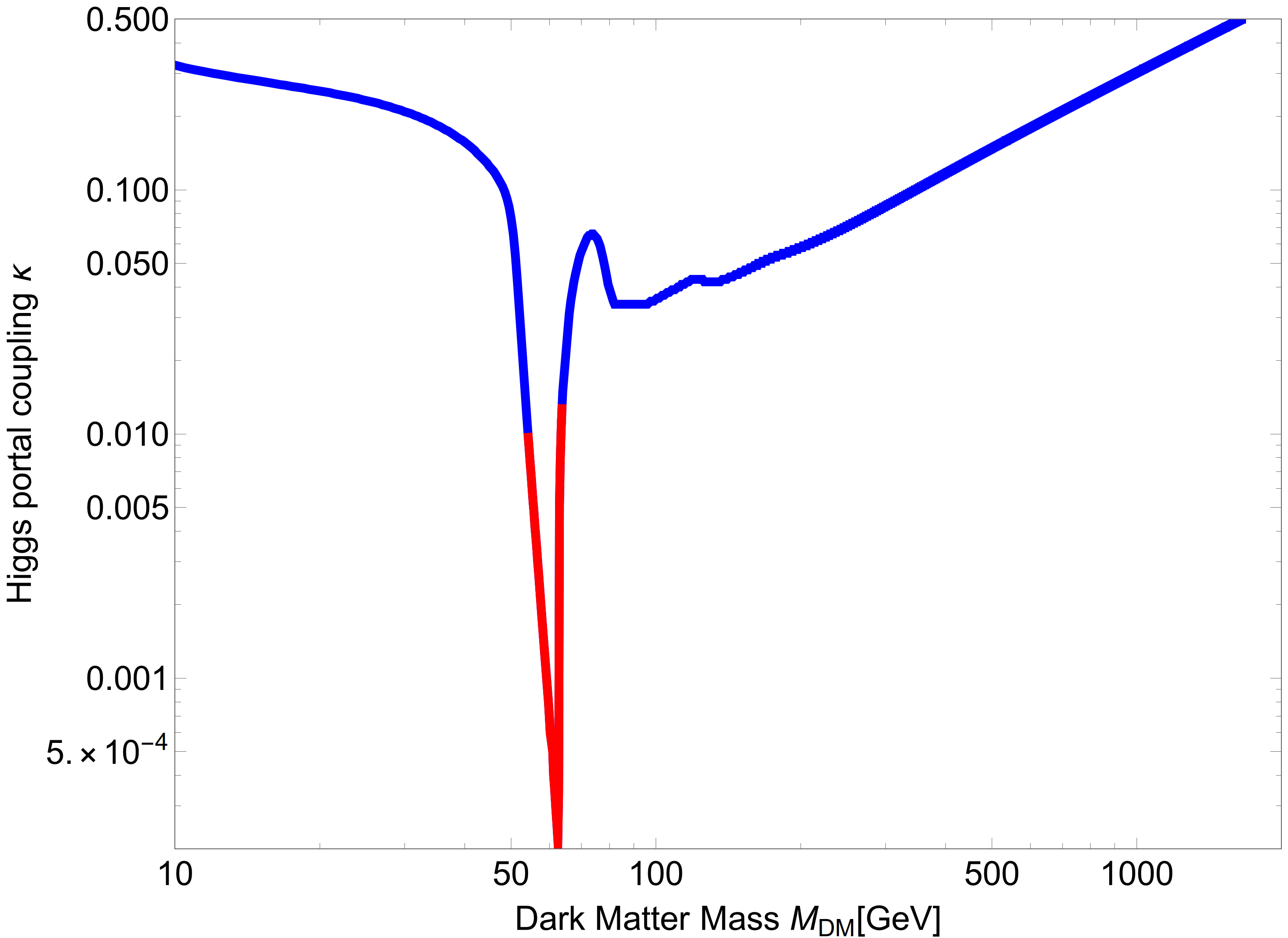}
		\hspace{0.2cm}
	\includegraphics[scale=.3]{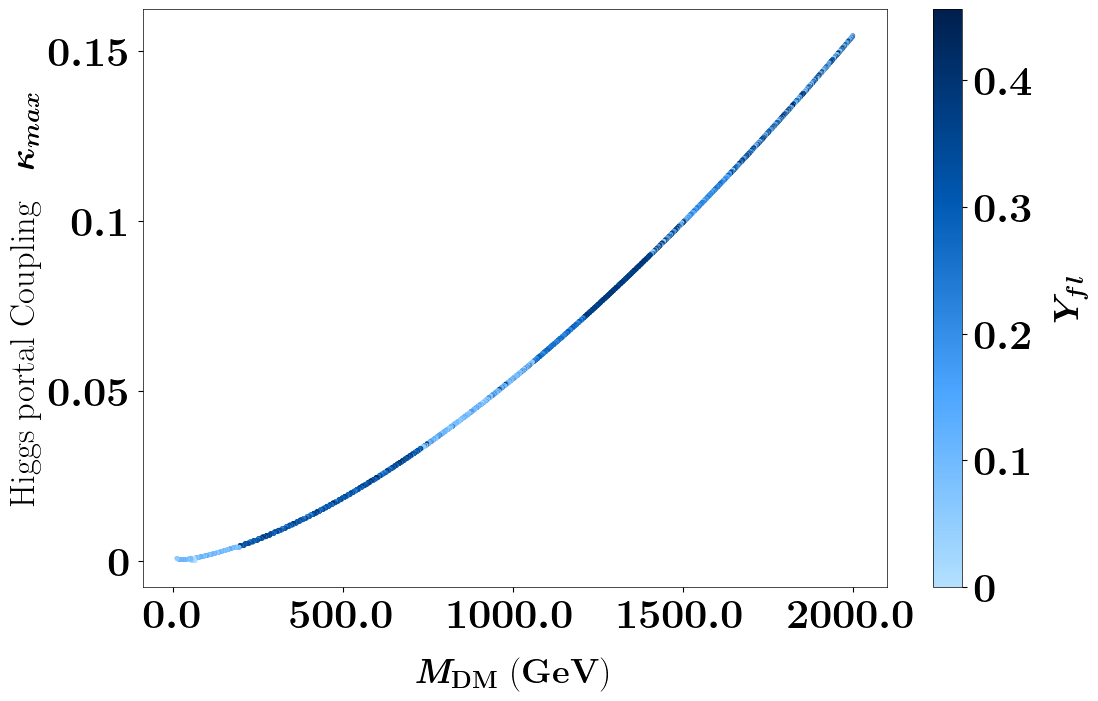}

	\caption{ \it The plots are obtained by varying the dark matter mass ($M_{\rm DM}$) and the Higgs portal coupling ($\kappa$). The blue line corresponds to the observed relic density, $\Omega_{\rm DM} h^2 = 0.1198$.In the left panel, the Yukawa coupling $Y_f=0$, so the effect of the new fermions on the relic density is absent. The red-blue bands give correct relic density, however red one is only allowed from the direct detection data. The right panel shows the maximum allowed $\kappa$ consistent with relic density, direct-detection, and other constraints. The corresponding values of $Y_f$ required to obtain the correct relic density are shown in the legend.}	
    \label{fig:relicDD}
\end{figure}

We first consider the minimal Higgs portal scenario in which the relic abundance is generated solely through the portal interaction $\kappa$. To suppress the effects of the additional fermions, we set $M_{E_1^\pm}=M_{E_2^\pm}=2.5~{\rm TeV}$,
and choose $Y_{fi}=0$. The mixing effects among the new fermions become negligible for this chosen benchmark values, which is leading to a suppressed contribution from $Y_N$. The corresponding parameter space consistent with the observed relic density is shown in the left panel of Fig.~\ref{fig:relicDD}, where the DM mass ($M_{\rm DM}$) and the Higgs portal coupling ($\kappa$) are varied.
\begin{table*}[h!]
	\centering
	\begin{tabular}{|p{1.5cm}|p{1.2cm}|p{1.4cm}|p{1.2cm}|p{1.2cm}|c|p{2.3cm}|p{4.7cm}|}
		\hline
		\hline
		Channel & $M_{DM}$ (GeV) & ~~$\kappa~~~$& $M_{E_1^\pm}$ (TeV) &~~$Y_{f}$&$\Omega_{DM}h^2$&DD-Cross section(cm$^2$)&~~~~~~~~~~Percentage \\
		\hline
		\hline				&&&&&&&$\sigma(s \,s\rightarrow b\bar{b})\quad~89\%$\\
		~~BP-1&30&0.215&2.5~~&0&0.1156& $4.59\times 10^{-43}$ &$\sigma(s \,s\rightarrow c\bar{c})\quad~7\%$\\
		&&&&&&&$\sigma(s \,s\rightarrow \tau^+\tau^-)\quad~5\%$\\
		\hline				&&&&&&&$\sigma(s \,s\rightarrow b\bar{b})\quad~76\%$\\
		&&&&&&&$\sigma(s \,s\rightarrow W^{\pm}W^\mp)\quad~13\%$\\
		~~BP-2&60&0.00071&2.5~~&0&0.1208& $1.15\times 10^{-48}$& $\sigma(s \,s\rightarrow  c\bar{c})\quad 6 \%$\\
		&&&&&&&$\sigma(s \,s\rightarrow \tau^-\tau^+)\quad~4\%$\\
		&&&&&&&$\sigma(s \,s\rightarrow ZZ)\quad~1\%$\\
		\hline
				&&&&&&&$\sigma(s \,s\rightarrow W^\pm W^\mp)\quad 46\%$ \\
		~~BP-3&500&0.145&2.5~~& 0 &0.1198& $1.00\times 10^{-45}$& $\sigma(s \,s\rightarrow HH) \quad23\%$\\
				&&&&&&&$\sigma(s \,s\rightarrow ZZ)\quad~23\%$\\
						&&&&&&&$\sigma(s \,s\rightarrow t\bar{t})\quad~7\%$\\
		\hline
				&&&&&&&$\sigma(s \,s\rightarrow W^\pm W^\mp)\quad 50\%$ \\
		~~BP-4&2000&0.62&2.5~~& 0 &0.1167& $8.32\times 10^{-46}$& $\sigma(s \,s\rightarrow HH) \quad 25\%$\\
				&&&&&&&$\sigma(s \,s\rightarrow ZZ)\quad~25\%$\\
		\hline
	\end{tabular}
	\caption{ \it Dark matter relic density is achieved through the $s$− and point-channel annihilation processes. Only a dark matter mass around 60 GeV remains allowed from the direct-detection (DD) cross-section and other phenomenological constraints.}
	\label{tabDM:1}
\end{table*}
The blue band represents the region satisfying the measured relic abundance, $\Omega_{\rm DM}h^2=0.1198\pm0.0026$,
while the red points denote the subset of parameter space that also survives current direct-detection limits~\cite{LZ:2024zvo}. The results indicate that, in the absence of additional interactions, a significant part of the singlet scalar dark matter parameter space is excluded by present direct-detection constraints. The benchmark points illustrating both allowed and excluded regions along with the dominant annihilation channels contributing to the relic density have been enlisted in Table~\ref{tabDM:1} . In this pure Higgs portal case, only the region near the Higgs resonance remains largely consistent with all existing bounds.

We next examine the impact of nonzero Yukawa interactions. The right panel of Fig.~\ref{fig:relicDD} shows the maximum allowed Higgs portal coupling consistent with relic density, direct-detection bounds, and other experimental constraints. The displayed parameter points satisfy
$\Omega_{\rm DM}h^2=0.1198\pm0.0026$
while remaining compatible with present limits from direct detection and Higgs invisible decay measurements. Representative benchmark points are summarized in Table~\ref{tabDM:2}.
\begin{table*}
	\centering
	\begin{tabular}{|p{1.5cm}|p{1.2cm}|p{1.4cm}|p{1.2cm}|p{1.2cm}|c|p{2.3cm}|p{4.7cm}|}
		\hline
		\hline
		Channel & $M_{DM}$ (GeV) & ~~$\kappa~~~$& $M_{E_1^\pm}$ (TeV) &~~$Y_{f}$&$\Omega_{DM}h^2$&DD-Cross section(cm$^2$)&~~~~~~~~~~Percentage \\
		\hline
		~~BP-5&30&0.0002&2.5& 0.365 &0.1157&$3.6\times 10^{-49}$ &$\sigma(ss\rightarrow \nu_\ell \bar{\nu}_\ell)\quad~100\%$ \\
		\hline
		~~BP-6&500&0.017&2.5& 0.375 &0.1157&$9.9\times 10^{-48}$ &$\sigma(ss\rightarrow \nu_\ell \bar{\nu}_\ell)\quad~100\%$ \\
		\hline
		~~BP-7&1000&0.05&2.5& 0.390 &0.1214&$2.16\times 10^{-47}$ &$\sigma(ss\rightarrow \nu_\ell \bar{\nu}_\ell)\quad~98\%$ \\
		&&&&&&& $\sigma(ss\rightarrow W^\pm W^\mp)\quad2\%$\\
		\hline
		&&&&&&&$\sigma(ss\rightarrow \nu_\ell \bar{\nu}_\ell)\quad~96\%$\\
		~~BP-8&1500&0.096&2.5& 0.423 &0.1193&$3.54\times 10^{-47}$ &$\sigma(ss\rightarrow W^\pm W^\mp)\quad2\%$ \\
		&&&&&&& $\sigma(ss\rightarrow Z Z)\quad1\%$\\
		&&&&&&& $\sigma(ss\rightarrow H H)\quad1\%$\\
		\hline
		&&&&&&&$\sigma(ss\rightarrow \nu_\ell \bar{\nu}_\ell)\quad~80\%$\\
		&&&&&&&$\sigma(E_1^\pm S\rightarrow W^{\pm} \nu_\ell)\quad 7\%$  \\
		~~BP-9&2000&0.14&2.5& 0.44 &0.1182&$4.24\times 10^{-47}$ &$\sigma(X_1S\rightarrow W^\pm l) \quad 6\%$ \\
		&&&&&&& $\sigma(ss\rightarrow W^\pm W^\mp)\quad3\%$\\
		&&&&&&& $\sigma(ss\rightarrow Z Z)\quad2\%$\\
		&&&&&&& $\sigma(ss\rightarrow H H)\quad2\%$\\
		\hline
	\end{tabular}
	\caption{ \it Dark matter relic density is achieved through the $s$− and point and $t,u$-channel annihilation processes. All these points are allowed from the direct-detection (DD) cross-section and other phenomenological constraints.}
	\label{tabDM:2}
\end{table*}
In this region of parameter space, the relic density is predominantly controlled by the $t$-channel annihilation process
$ss\rightarrow \nu_\ell\bar{\nu}_\ell$,
because the Higgs portal contribution is strongly restricted by direct-detection data. As a result, the Yukawa interactions play the leading role in obtaining the observed relic abundance. For heavier DM masses, particularly around $M_{\rm DM}\sim2~{\rm TeV}$,
co-annihilation channels involving the additional $Z_2$-odd fermions become increasingly important due to the small mass splitting between the DM particle and the next-to-lightest states. It should be noted that the contribution from the interaction term 
$Y_{fi}^{\prime }\, \overline{l}_{i,R} E_S \,s$ remains highly suppressed for $Y_{fi}^{\prime}=\mathcal{O}(0.1)$, due to the heavy charged fermion mass $M_{E_{2}^{\pm}} = 3.0$ TeV together with $\cos\beta = 0.995$. Therefore, for simplicity, we neglect the effect of this interaction term (see eqn.~\ref{lint}) by setting $Y_{fi}^{\prime}=0$ in this analysis.

\begin{figure}
	\centering
    	\includegraphics[scale=.29]{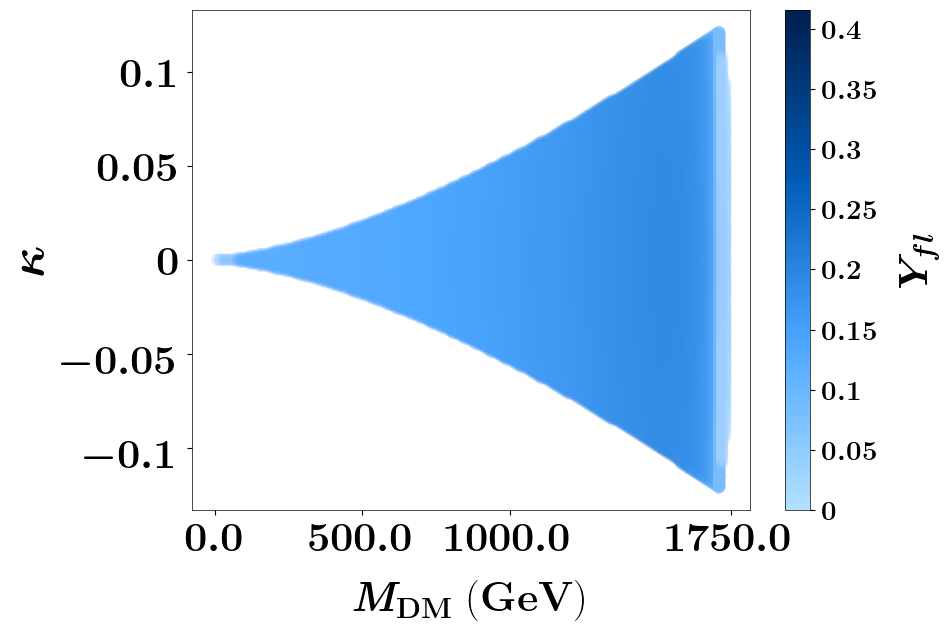}
		\hspace{0.4cm}
	\includegraphics[scale=.4]{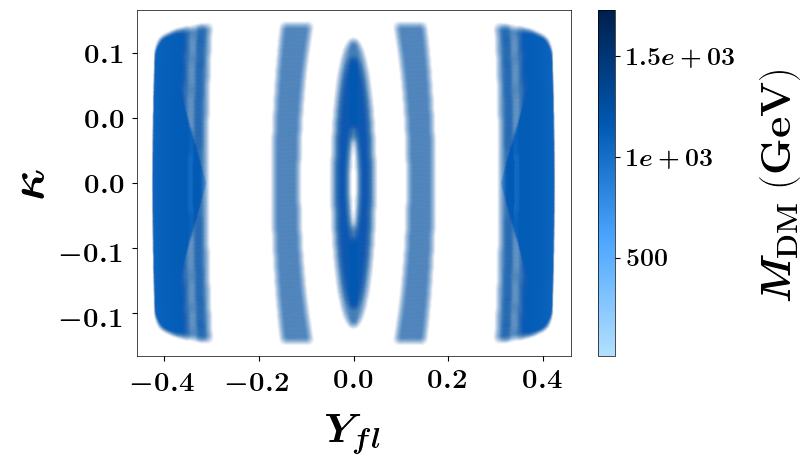}\\
	\includegraphics[scale=.29]{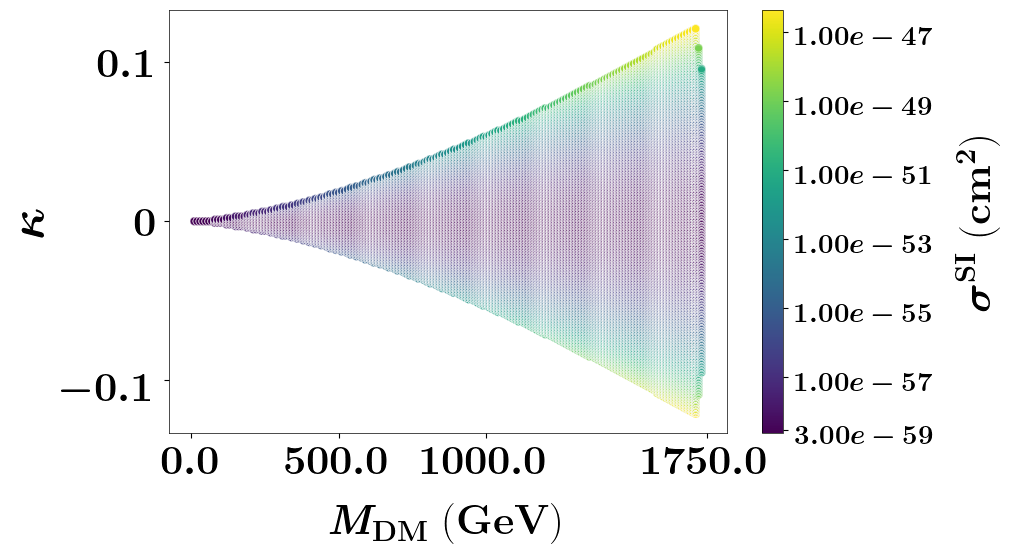}
	\includegraphics[scale=.35]{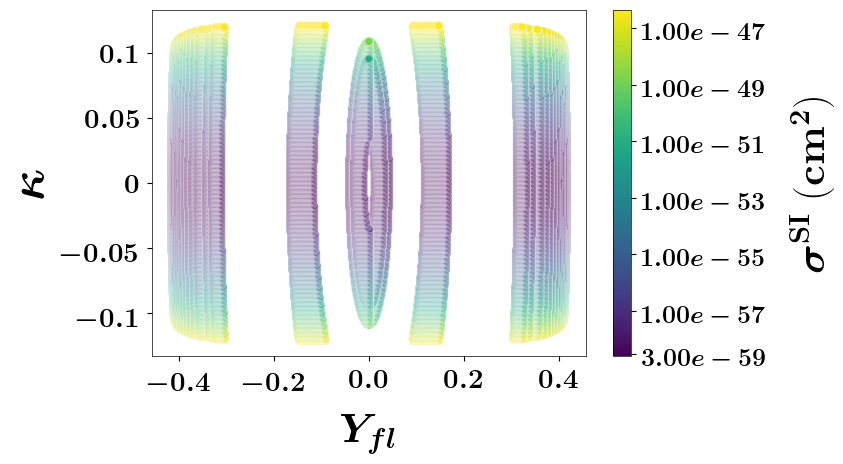}
	\caption{ \it The coloured band represents the relic density within the $3\sigma$ interval, $0.1120 <$ $ \Omega_{\rm DM} h^2 $ $< 0.1276$. The plots are obtained by varying the dark matter mass, the Higgs portal coupling $\kappa$, and the new Yukawa coupling $Y_f$. The colour gradients in the upper panel represent the variation of the Yukawa coupling and the dark matter mass, respectively. The colour gradient in the lower panel corresponds to the direct-detection cross-section values in the same plot. The blank regions are excluded by at least one of the following constraints: vacuum stability, unitarity, direct detection bounds, or the relic density requirement $0.1120 <$ $ \Omega_{\rm DM} h^2 $ $< 0.1276$. The parameter space consistent with the observed relic density also satisfies the current limits from lepton flavour violating (LFV) processes, as well as the anomalous magnetic moments of the electron and muon.}	\label{fig:relicall}
\end{figure}
We perform a comprehensive scan over the DM mass ($M_{\rm DM}$), the Higgs portal coupling ($\kappa$), and the new Yukawa coupling ($Y_f$) to explore the full allowed parameter space of the model. The dark matter mass $M_{\rm DM}$ is varied from approximately $5$ GeV to $2.0$ TeV in steps of $5$ GeV. The Higgs portal coupling $\kappa$ is scanned in the range $0 \leq \kappa \leq 0.15$ with a step size of $0.0001$, since values beyond $\kappa = 0.15$ are already excluded by direct-detection constraints for $M_{\rm DM}=2.0$ TeV. The new Yukawa coupling $Y_f$ is varied from $0$ to $0.5$ with a step size of $0.0005$. The corresponding allowed parameter regions are shown in the $\kappa-M_{\rm DM}$ plane in Fig.~\ref{fig:relicall}(left: top and bottom) and in the $\kappa-Y_f$ plane in Fig.~\ref{fig:relicall}(right: top and bottom). In this section, we keep fix the charged fermion masses at $M_{E_1^{\pm}} = 2$ TeV and $M_{E_2^{\pm}} = 3$ TeV, which remain consistent with the current LHC constraints and the projected sensitivities of future HL-LHC and muon collider searches. All coloured bands in Fig.~\ref{fig:relicall} correspond to the DM relic density within the $3\sigma$ interval, $0.1120 < \Omega_{\rm DM} h^2 < 0.1276$. The colour distribution in the left plot of the upper panel illustrates the dependence on the Yukawa coupling. In contrast, the dark matter mass variation is shown in the right plot of the upper panel.
In the lower panel, both left and right plots, the colour scale indicates the corresponding values of the direct-detection cross-section. The empty regions are ruled out by inclusive or the following conditions: vacuum stability, perturbative unitarity, direct-detection limits, or the relic density constraint $0.1120 < \Omega_{\rm DM} h^2 < 0.1276$. Furthermore, the parameter space compatible with the observed relic abundance remains consistent with existing bounds from lepton flavour violation (LFV) searches and the anomalous magnetic moments of both the electron and the muon.

\begin{figure}[h!]
	\centering
    	\includegraphics[scale=0.35]{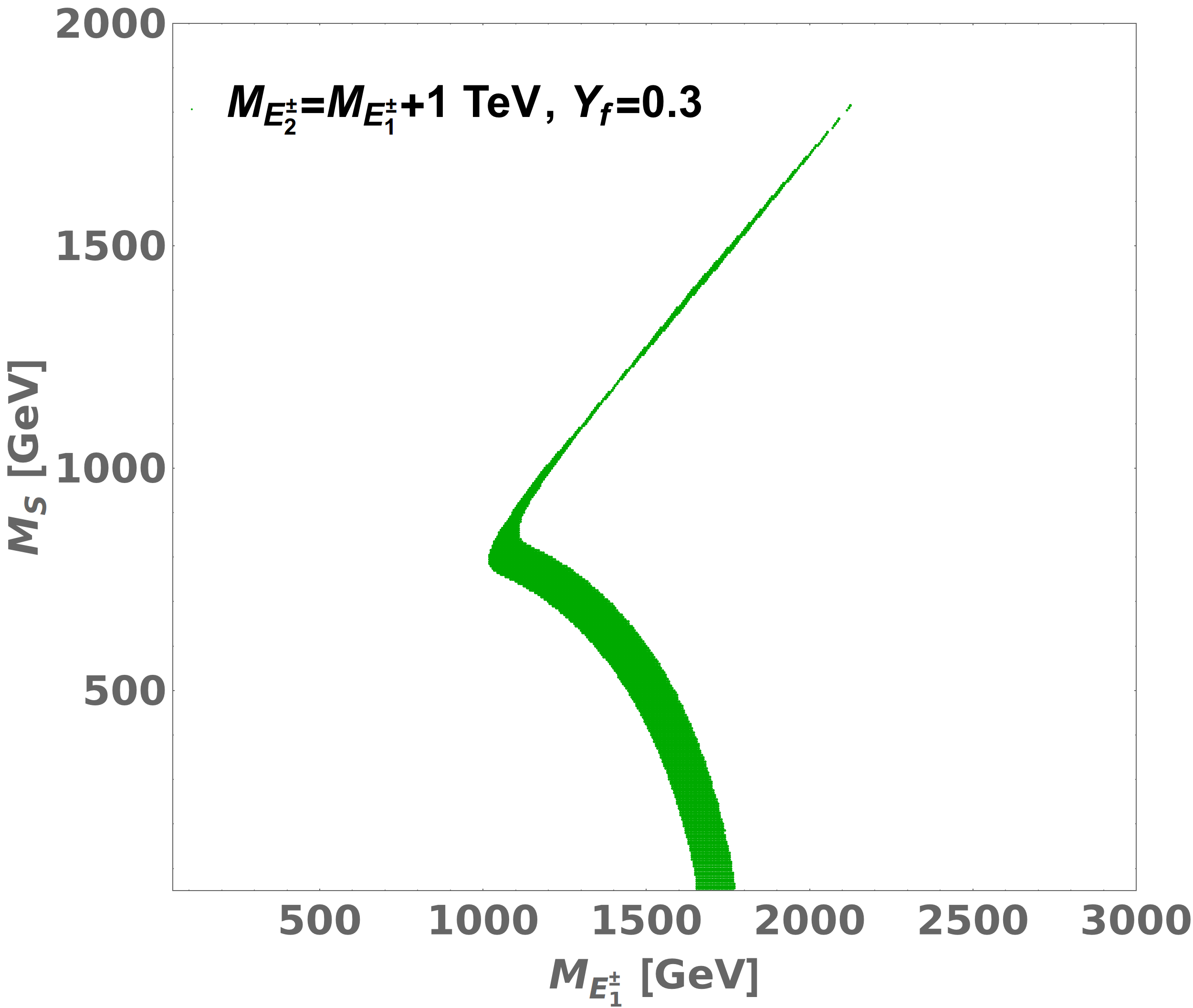}
    	\includegraphics[scale=0.35]{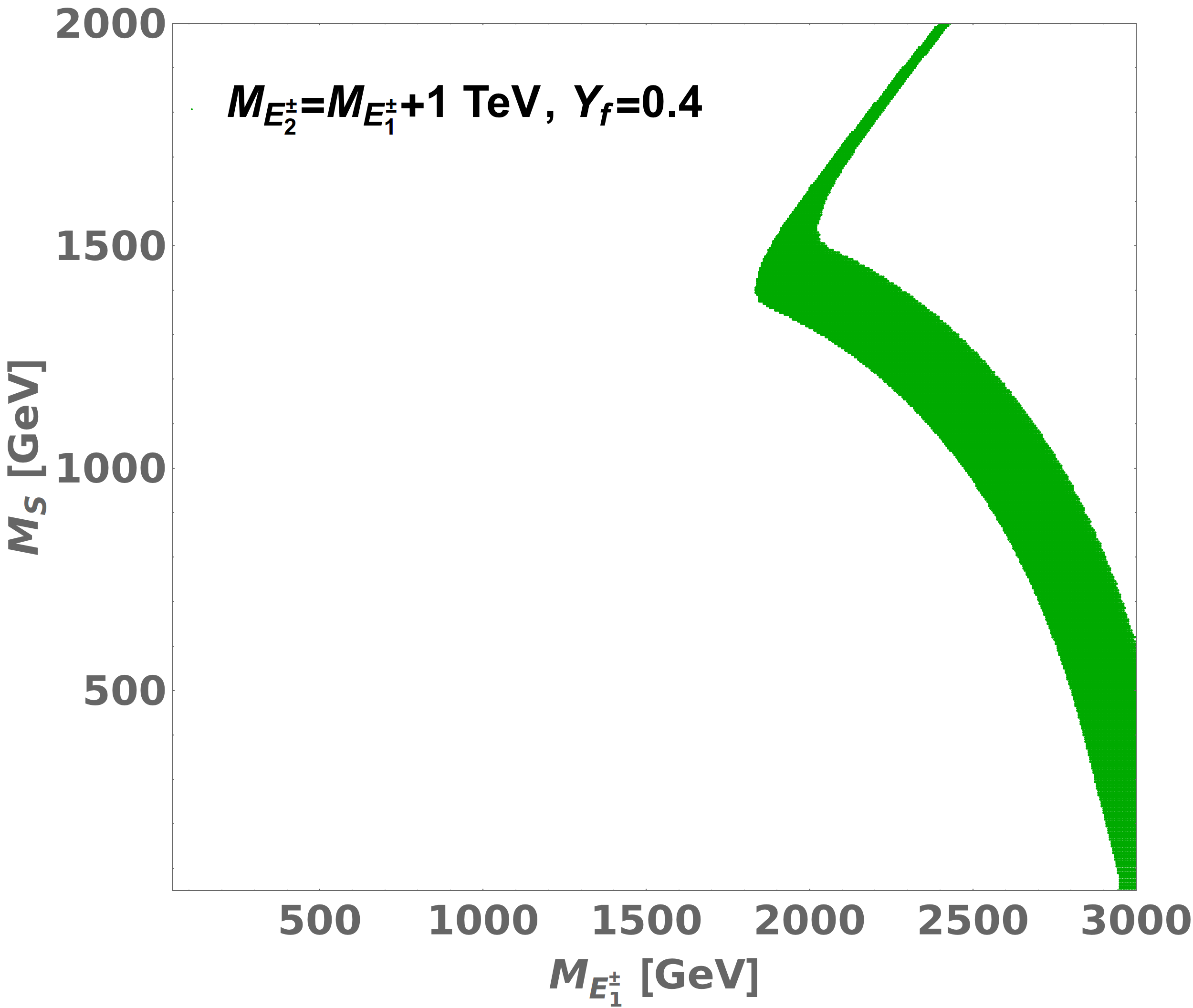}
	\caption{  \it The coloured band indicates the relic density within the $3\sigma$ range, $0.1120 <$ $ \Omega_{\rm DM} h^2 $ $< 0.1276$. The plots are obtained by varying the dark matter mass and the charged fermion mass $(M_{E^\pm})$, while fixing the Yukawa coupling at $Y_f = 0.3 ~(left)$ and $Y_f = 0.3~ (right)$. This allowed region is consistent with the relic density, direct-detection limits, and other experimental constraints.}	
    \label{fig:relicME1}
\end{figure}
In Fig.~\ref{fig:relicME1}, we show the variation of the lightest charged fermion mass $M_{E_1^\pm}$ as a function of the DM mass. The new Yukawa coupling is fixed at $Y_f=0.3$ and $0.4$ respectively. For this analysis, the DM mass is scanned from $10$ GeV to $2.0$ TeV with a step size of $5$ GeV, while $M_{E_1^\pm}$ is varied between $200$ GeV and $2.0$ TeV with the same step size. The heavier charged fermion mass is taken as $M_{E_2^\pm}=M_{E_1^\pm}+1.0~\text{TeV}$. In addition, the Higgs portal coupling ($\kappa$) is varied within the limits allowed by direct-detection constraints for each DM mass point in order to identify a broad parameter region consistent with the observed relic density. This figure is particularly useful for understanding the parameter space relevant to the future muon collider study discussed in the collider section.

\section{Collider Searches}
\label{sec:col}
In this section, we present a collider study of the lightest charged fermion, $E_1^\pm$, at future muon colliders for center-of-mass energies, $\sqrt{s} = 3$~TeV (with projected integrated luminosity, $\mathcal{L}=1~\text{ab}^{-1}$) and $\sqrt{s} = 10$~TeV (projected $\mathcal{L}=10~\text{ab}^{-1}$)~\cite{InternationalMuonCollider:2025sys}.
\subsection{Signal Process}
\begin{figure}[htbp]
\centering

\begin{minipage}[t]{0.48\textwidth}
\centering
\begin{tikzpicture}
\begin{feynman}

\vertex (i1) at (-2,  1.5) {$\mu^-$};
\vertex (i2) at (-2, -1.5) {$\mu^+$};

\vertex (a) at (0,0);
\vertex (b) at (2,0);

\vertex (c1) at (4,  1.2);
\vertex (c2) at (4, -1.2);

\vertex (f1) at (6,  2.0) {$e^-$};
\vertex (s1) at (6,  0.4) {$S$};
\vertex (f2) at (6, -0.4) {$e^+$};
\vertex (s2) at (6, -2.0) {$S$};

\diagram*{
  (i1) -- [fermion] (a),
  (i2) -- [anti fermion] (a),
  (a) -- [boson, edge label={$\gamma/ Z/ h$}] (b),
  (b) -- [fermion, edge label=$E_1^-$] (c1),
  (b) -- [anti fermion, edge label'=$E_1^+$] (c2),
  (c1) -- [fermion] (f1),
  (c1) -- [scalar] (s1),
  (c2) -- [anti fermion] (f2),
  (c2) -- [scalar] (s2),
};

\end{feynman}
\end{tikzpicture}

\vspace{0.3cm}
(a) $s$-channel
\end{minipage}
\hfill
\begin{minipage}[t]{0.48\textwidth}
\centering
\begin{tikzpicture}
\begin{feynman}

\vertex (i1) at (-3,  1.5) {$\mu^-$};
\vertex (i2) at (-3, -1.5) {$\mu^+$};

\vertex (a) at (-1,  1.0);
\vertex (b) at (-1, -1.0);

\vertex (o1) at (2,  1.5); 
\vertex (o2) at (2, -1.5); 

\vertex (f1) at (5,  2.2) {$e^-$};
\vertex (s1) at (5,  0.8) {$S$};
\vertex (f2) at (5, -0.8) {$e^+$};
\vertex (s2) at (5, -2.2) {$S$};

\diagram*{
  (i1) -- [fermion] (a) -- [fermion, edge label={$E_1^-$}] (o1),
  (i2) -- [anti fermion] (b) -- [anti fermion, edge label'={$E_1^+$}] (o2),
  (a) -- [scalar, edge label'={$S$}] (b),

  (o1) -- [fermion] (f1),
  (o1) -- [scalar] (s1),
  (o2) -- [anti fermion] (f2),
  (o2) -- [scalar] (s2),
};

\end{feynman}
\end{tikzpicture}

\vspace{0.3cm}
(b) $t$-channel
\end{minipage}

\caption{ \it Production of lightest charged fermions $E_1^+ \: \textrm{and} \: E_1^-$ in $\mu^+\mu^-$ collisions through the $s$-channel (a) and $t$-channel (b) processes, followed by the decays of $E_1^\pm \to e^\pm S$.}
\label{fig:feynman_diagram}
\end{figure}
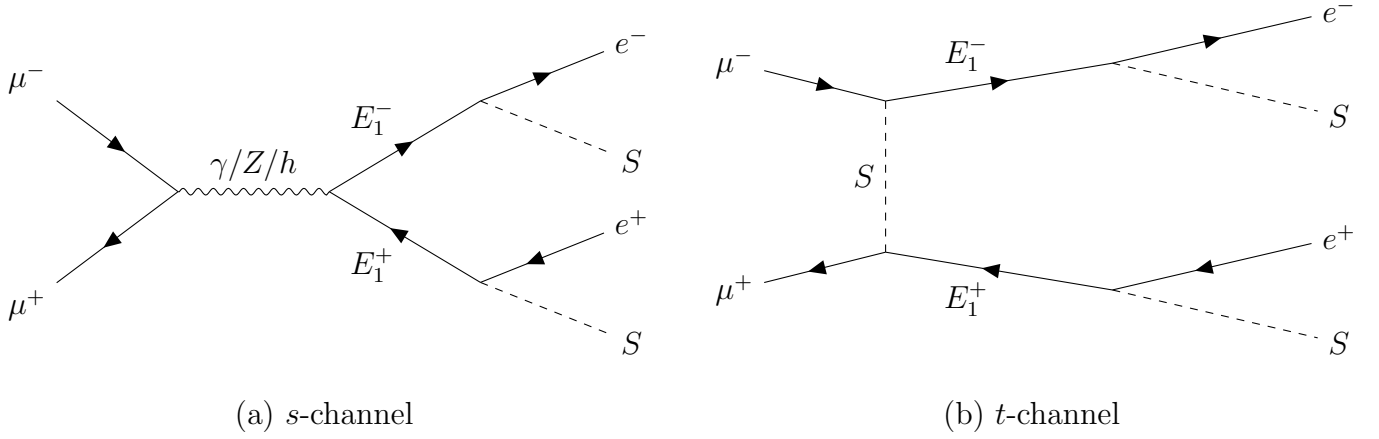

We consider the signal process $$\mu^+\mu^- \rightarrow E_1^+ E_1^-$$ where $E_1^\pm$ subsequently decays into an electron and the lightest stable particle (LSP), DM, i.e., $E_1^\pm \rightarrow e^\pm S$. We focus on the electron channel for the decay of $E_1^\pm$. The muon channel is not considered because the corresponding interactions are strongly constrained by measurements of the muon anomalous magnetic moment, $(g-2)_\mu$ and $\mu\to e \gamma$~\cite{Kawamura:2020qxo, Yin:2021yqy}. Restricting the analysis to the electron channel allows us to avoid these stringent low-energy constraints and concentrate on the collider phenomenology of the model. The Feynman diagram corresponding to the signal process is shown in Fig.~\ref{fig:feynman_diagram}, where the production of $E_1^\pm$ proceeds via both $s$-channel and $t$-channel exchanges. At the detector level, the final state is characterized by two oppositely charged electrons and significant missing transverse energy, $\cancel{E}_T$, arising from the undetected LSPs.

\subsection{Standard Model Backgrounds}

The primary Standard Model (SM) backgrounds can mimic this signature either through the presence of genuine neutrinos in the final state (irreducible backgrounds) or through detector effects such as mis-measurement or object misidentification (reducible backgrounds). The background contributions considered in this analysis include the following SM processes.
\begin{enumerate}
    \item $\mu^+\mu^- \to e^+ e^- \nu_\ell \bar{\nu}_\ell$ [Bkgd I] 
    \item $\mu^+\mu^- \to e^+ e^-, \;e^+ e^-jj,\;e^+ e^-\gamma\gamma$ [Bkgd II] 
    \item $\mu^+ \mu^- \to W^+(\to e^+ \nu_e)\; W^-(\to e^-\bar{\nu}_e)\; \gamma$ [Bkgd III] 
    \item $\mu^+ \mu^- \to \tau^+ (\to e^+\nu_e\bar{\nu}_\tau)\;\tau^-(\to e^+\bar{\nu}_e\nu_\tau)$ [Bkgd IV]
\end{enumerate}
The dominant irreducible background arises from Bkgd I, where the neutrinos escape detection and give rise to genuine $\cancel{E_T}$. This process closely resembles the signal topology, making it the most challenging background to suppress. Bkgd II does not contain intrinsic sources of missing energy at the parton level. However, detector effects such as imperfect energy reconstruction and particles escaping through the forward acceptance of the detector can induce spurious $\cancel{E}_T$, allowing a non-negligible fraction of these events to mimic the signal signature. A subleading but important irreducible background comes from Bkgd III. In this case, the leptonic decays of the $W$ bosons produce electrons and neutrinos, leading to a final state with two electrons and genuine $\cancel{E_T}$. The additional photon may be soft or emitted in the forward region, and hence can evade detection, making the event topology similar to the signal. And, finally, Bkgd IV constitutes a suppressed but non-negligible background where, both taus decay leptonically into electrons and multiple neutrinos, resulting in a final state with two electrons and substantial missing energy. We will show the projected exclusion/discovery reach of direct heavy charged fermion and dark matter as transverse missing energy searches in this channels by performing a detailed cut based collider analysis.

\subsection{Event Generation}

\begin{figure}[t]
    \centering
    \includegraphics[width=0.7\textwidth]{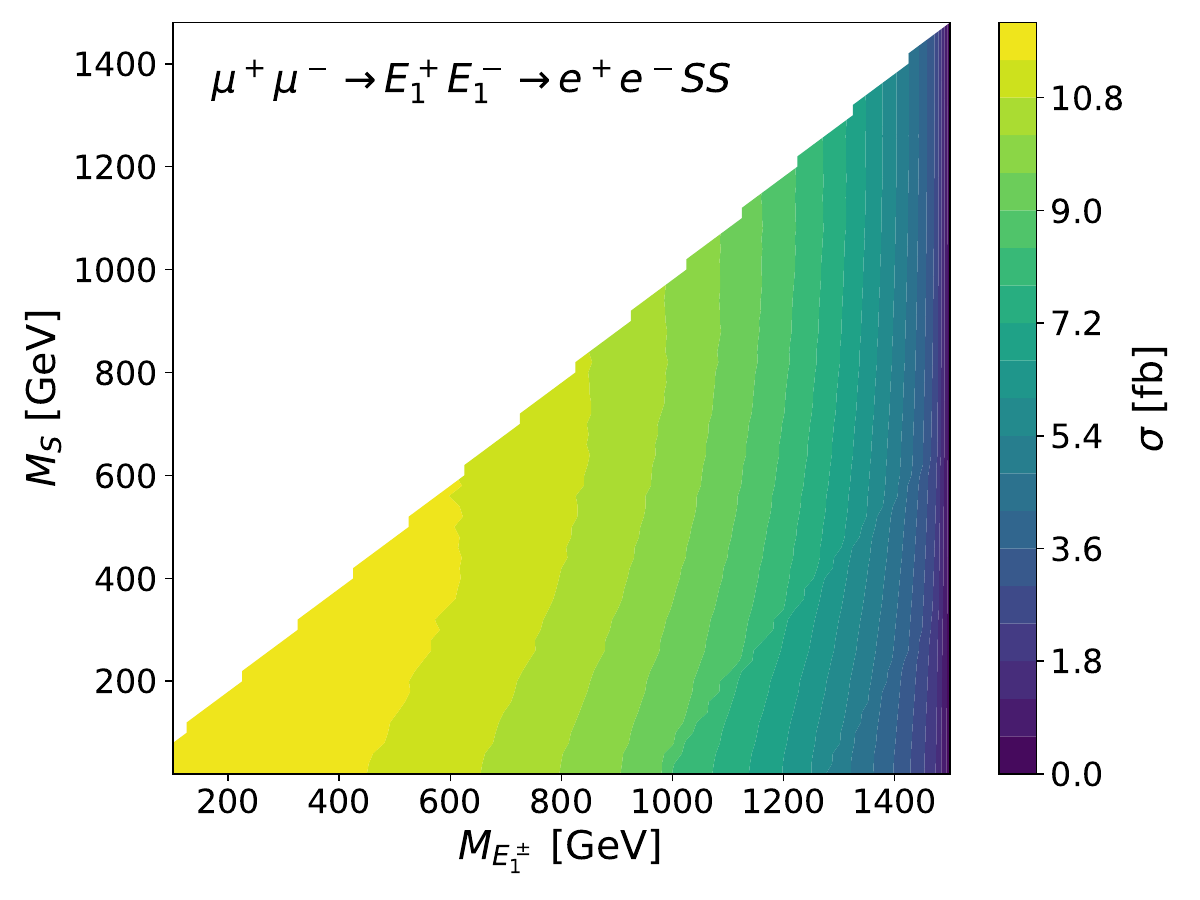}
    \caption{ \it Contour plot of the signal cross section for $\mu^+\mu^- \to E_1^+E_1^- \to e^+e^-SS$ in the $(M_{E_1^\pm},\, M_S)$ plane at $\sqrt{s}=3$~TeV. The colour bar indicates the cross section $\sigma$ in fb. The shaded kinematic region corresponds to the parameter space where the decay $E_1^\pm \to e^\pm S$ is allowed.}
    \label{fig:xsec_ME1_MS}
\end{figure}

We begin by simulating both the signal and background processes using a multi-stage framework that models the full evolution from parton-level event generation to detector-level reconstruction. For the signal process, the model is first implemented in \texttt{FeynRules 2.0}~\cite{Alloul:2013bka}, from which the Universal FeynRules Output (UFO)~\cite{Degrande:2011ua} files are generated. These UFO files are then used within \texttt{MadGraph5\_aMC@NLO}~\cite{Alwall:2014hca} to simulate the hard scattering events, $$\mu^+ \mu^- \to E_1^+ E_1^-,\: E_1^+ \to e^+ S,\: E_1^- \to e^- S$$ at $\sqrt{s} = 3$~\text{TeV}($\mathcal{L}=1$~ab$^{-1}$)~\footnote{We repeat the collider analysis for $\sqrt{s}=10$~TeV ($\mathcal{L}=10$~ab$^{-1}$), using the same signal and background processes as in the $\sqrt{s}=3$~TeV case. The event generation and simulation framework remain unchanged.}. Since we are looking into the electron channel of $E_1^\pm$, we kept the parameter $Y_{f1} = 0.3$ and $Y_{f2,f3}\sim 0$. No additional generator level cuts are imposed at this stage. The generated events are subsequently interfaced with \texttt{PYTHIA~8.3}~\cite{Sjostrand:2006za, SJOSTRAND2008852}, which performs parton showering and hadronization. Finally, the events are passed through \texttt{Delphes~3.5.0}~\cite{deFavereau:2013fsa}, employing the default muon collider detector card to simulate the detector response. 10000 signal events are generated for each point in the $(M_{E_1^\pm}, M_S)$ parameter space. The charged fermion mass $M_{E_1^\pm}$ is varied from 100~GeV to 1500~GeV in steps of 25~GeV, while the dark matter mass $M_S$ is varied from 20~GeV up to values below $M_{E_1^\pm}$ in steps of 20~GeV~\footnote{Note that, in order to satisfy the constraints from the electroweak precision parameters $S$, $T$, and $U$, we take $\cos\beta = 0.995$.}. The resulting signal cross section, $\sigma$, in the two-dimensional $(M_{E_1^\pm},\, M_S)$ plane is shown in Fig.~\ref{fig:xsec_ME1_MS}.

From the plot, one observes that the cross section is largest in the region of low $M_{E_1^\pm}$ and decreases steadily as the mass of $E_1^\pm$ increases. This behavior is expected due to phase-space suppression and the reduced production rate of heavier particles. Additionally, for a fixed $M_{E_1^\pm}$, the cross section shows a mild dependence on $M_S$. The accessible parameter space is bounded by the kinematic condition $M_S < M_{E_1^\pm}$, beyond which the decay $E_1^\pm \to e^\pm S$ is not allowed. Motivated by this behavior, we select the following benchmark points (BPs) tabulated in Table~\ref{tab:benchmarks} which are eventually used to design and optimize kinematical cuts for suppressing the SM backgrounds.

\begin{table}[h]
\centering
\begin{tabular}{| c | c | c | c |}
\hline
Benchmark Point & $M_{E_1^\pm}$ (GeV) & $M_S$ (GeV) & $\sigma$ (fb) \\
\hline
BP1 & 100 & 20 & 11.82 \\
BP2 & 100 & 40 & 11.81 \\
BP3 & 100 & 80 & 11.86 \\ \hline
BP4 & 800 & 20 & 10.18 \\
BP5 & 800 & 400 & 10.78 \\
BP6 & 800 & 700 & 10.95 \\ \hline
BP7 & 1400 & 20 & 3.57 \\
BP8 & 1400 & 700 & 5.15 \\
BP9 & 1400 & 1200 & 5.49 \\
\hline
\end{tabular}
\caption{ \it Benchmark points with corresponding masses and cross sections.}
\label{tab:benchmarks}
\end{table}

The SM backgrounds are simulated using the same framework as the signal, employing \\ \texttt{MadGraph5\_aMC@NLO} for event generation, followed by \texttt{PYTHIA~8.3} for parton showering and hadronization, and \texttt{Delphes~3.5.0} for detector simulation. The corresponding cross sections (in fb) and the number of generated events at $\sqrt{s}=3$~TeV ($\mathcal{L}=1$~ab$^{-1}$) for each background are summarized in Table~\ref{tab:backgrounds}.

\begin{table}[h]
\centering
\begin{tabular}{| c | c | c |}
\hline
Backgrounds & $\sigma$ (fb) & Events Generated \\
\hline
Bkgd I & 63.6 & 70000 \\
Bkgd II & 12.24 & 15000 \\
Bkgd III & 0.35 & 5000 \\
Bkgd IV & 0.005 & 10000 \\
\hline
\end{tabular}
\caption{S \it M background processes with their corresponding cross sections and number of generated events at $\sqrt{s}=3$~TeV ($\mathcal{L}=1$~ab$^{-1}$).}
\label{tab:backgrounds}
\end{table}

\subsection{Kinematic Distributions and Selection Strategy}

\begin{figure}
    \centering
    \includegraphics[width=\textwidth]{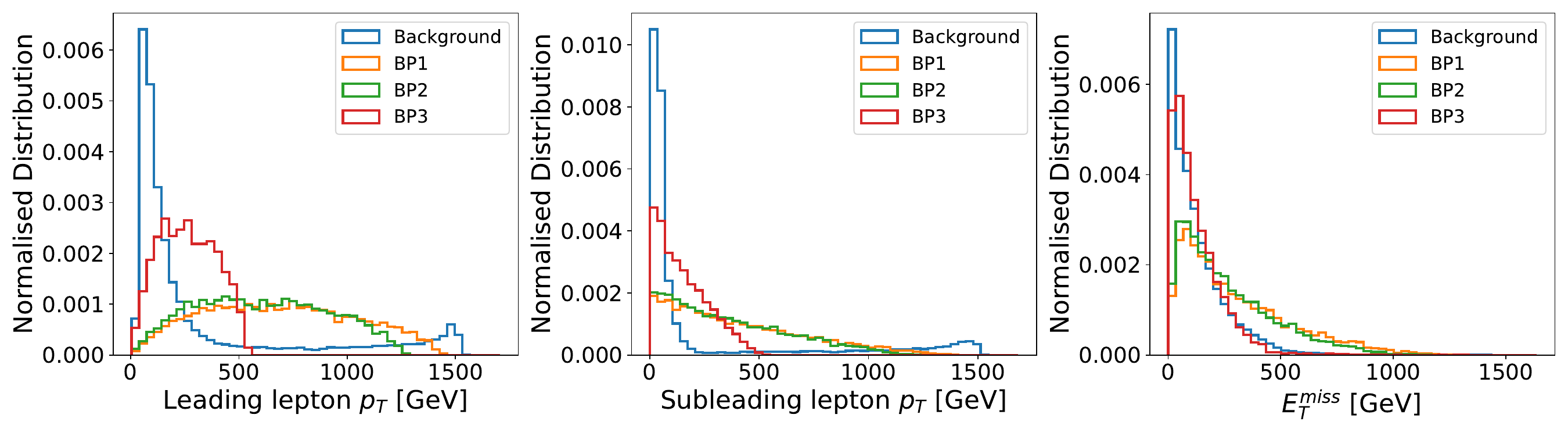}
    \caption{ \it Normalized distributions of leading lepton $p_T$, subleading lepton $p_T$, and missing transverse energy $\cancel{E_T}$ for benchmark points BP1--BP3 compared with the SM backgrounds, for $M_{E_1^\pm} = 100$~GeV. All the 4 backgrounds are combined according to their respective weights.}
    \label{fig:pt_met_ME100}
\end{figure}

\begin{figure}
    \centering
    \includegraphics[width=\textwidth]{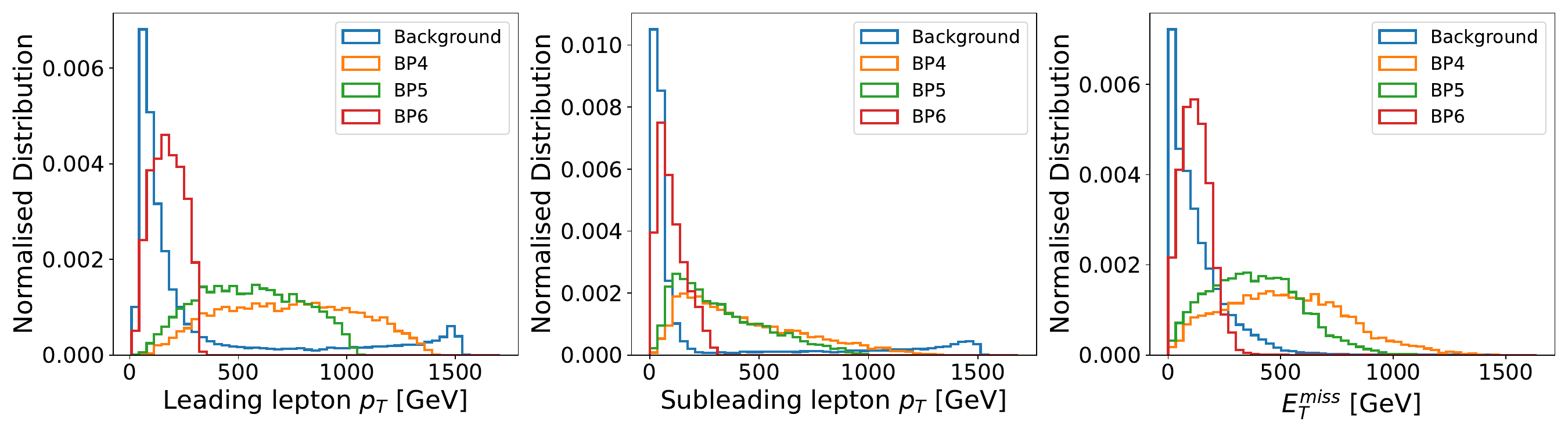}
    \caption{ \it Same as Fig.~\ref{fig:pt_met_ME100}, but for $M_{E_1^\pm} = 800$~GeV (BP4--BP6).}
    \label{fig:pt_met_ME800}
\end{figure}

\begin{figure}
    \centering
    \includegraphics[width=\textwidth]{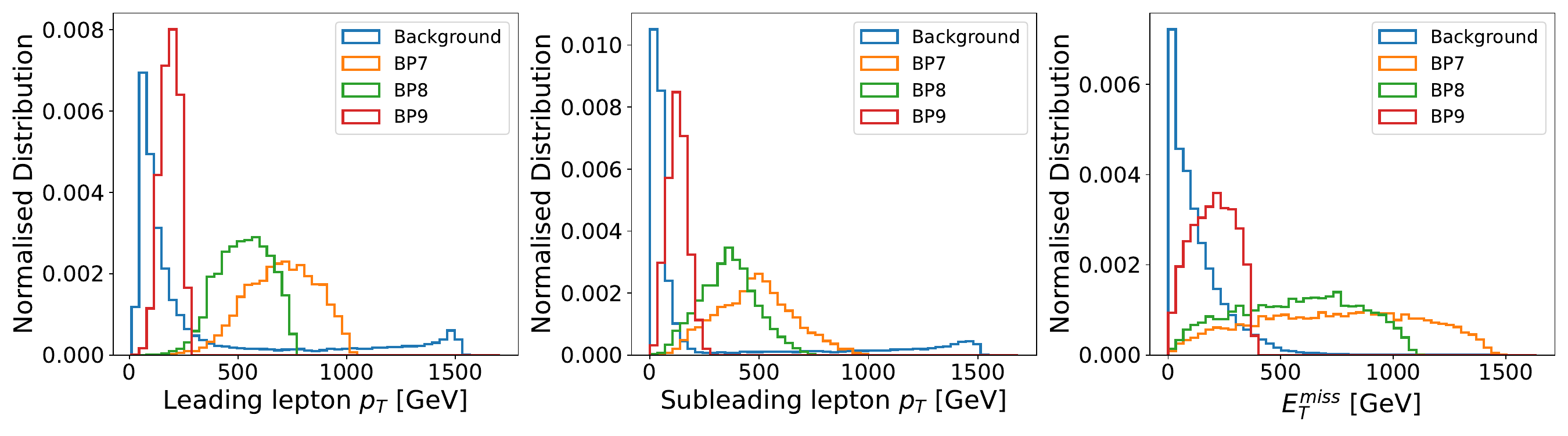}
    \caption{ \it Same as Fig.~\ref{fig:pt_met_ME100}, but for $M_{E_1^\pm} = 1400$~GeV (BP7--BP9).}
    \label{fig:pt_met_ME1400}
\end{figure}

Following the event generation and detector simulation, we perform a detailed event-level analysis on the benchmark points listed in Table~\ref{tab:benchmarks} to construct kinematic observables relevant for signal discrimination. Events are required to contain at least two reconstructed oppositely charged electrons and non-zero missing transverse energy. For events passing these selection criteria, the transverse momentum ($p_T$) distributions of the two highest-$p_T$ oppositely charged electrons ($p_T^{\ell_1}$ and $p_T^{\ell_2}$), along with the missing transverse energy ($\cancel{E_T}$) distribution, are shown for all nine benchmark points, together with the Standard Model backgrounds, in Figs.~\ref{fig:pt_met_ME100}--\ref{fig:pt_met_ME1400}. All events are appropriately weighted to reproduce the expected event yields at $\sqrt{s}=3$~TeV. 

From the aforementioned figures, we define three categories of kinematical selection cuts aimed at enhancing the signal significance across the nine benchmark points:
\begin{itemize}
    \item ``Loose'' cut : $p_T^{\ell_1} > 200$~GeV, $p_T^{\ell_2} > 100$~GeV, $\cancel{E}_T > 100$~GeV 
    \item ``Medium'' cut : $p_T^{\ell_1} > 250$~GeV, $p_T^{\ell_2} > 150$~GeV, $\cancel{E}_T > 200$~GeV
    \item ``Hard'' cut : $p_T^{\ell_1} > 300$~GeV, $p_T^{\ell_2} > 200$~GeV, $\cancel{E}_T > 300$~GeV   
\end{itemize}

\subsection{Signal Significance}

To evaluate the signal significance $\mathcal{S}_{\text{sig}}$ for different benchmark points, we employ the following expression given in Eq.~\ref{eq:significance}:
\begin{equation}
\mathcal{S}_{\text{sig}} = \sqrt{2 \left[
(S + B)\ln \left( 1 + \frac{S}{B + \epsilon^2 B (S + B)} \right)
- \frac{1}{\epsilon^2}\ln \left( 1 + \frac{\epsilon^2 S}{1 + \epsilon^2 B} \right)
\right]}
\label{eq:significance}
\end{equation}
where $S$ denotes the number of signal events surviving the selection cuts, $B$ denotes the number of background events surviving the selection, and $\epsilon$ represents the fractional systematic uncertainty on the background, which is taken to be $10\%$.\footnote{We adopt a conservative estimate for the systematic uncertainty, as the experimental conditions for a future muon collider are not yet fully established.} The resulting signal and background yields, together with the corresponding signal significances obtained after applying the three categories of selection cuts, are summarized in Table~\ref{tab:significance_table}.

\begin{table}[htbp]
\centering
\renewcommand{\arraystretch}{1.3}
\setlength{\tabcolsep}{5pt}
\resizebox{\textwidth}{!}{
\begin{tabular}{|c|ccc|ccc|ccc|}
\hline
\multirow{2}{*}{\textbf{Benchmark Points}} 
& \multicolumn{3}{c|}{\textbf{Loose}} 
& \multicolumn{3}{c|}{\textbf{Medium}} 
& \multicolumn{3}{c|}{\textbf{Hard}} \\
\cline{2-10}
& $S$ & $B$ & $\mathcal{S}_{\rm sig}$ 
& $S$ & $B$ & $\mathcal{S}_{\rm sig}$ 
& $S$ & $B$ & $\mathcal{S}_{\rm sig}$ \\
\hline
BP1 & 4447 & 7712 & 13.98 & 2807 & 3714 & 22.45 & 1753 & 1923 & 27.00 \\ \hline
BP2 & 4209 & 7712 & 13.42 & 2527 & 3714 & 20.91 & 1411 & 1923 & 23.43 \\ \hline
BP3 & 1745 & 7712 & 6.73  & 369  & 3714 & 4.72  & 20   & 1923 & 0.62  \\ \hline
BP4 & 6389 & 7712 & 16.30 & 5235 & 3714 & 30.48 & 4080 & 1923 & 41.40 \\ \hline
BP5 & 5754 & 7712 & 15.76 & 4187 & 3714 & 27.44 & 2746 & 1923 & 33.76 \\ \hline
BP6 & 1097 & 7712 & 4.21  & 8    & 3714 & 0.11  & 0    & 1923 & 0.00  \\ \hline
BP7 & 6953 & 7712 & 7.82  & 6523 & 3714 & 17.63 & 5892 & 1923 & 27.26 \\ \hline
BP8 & 6800 & 7712 & 10.29 & 5995 & 3714 & 21.40 & 5073 & 1923 & 31.26 \\ \hline
BP9 & 2030 & 7712 & 3.95  & 86   & 3714 & 0.58  & 0    & 1923 & 0.00  \\ \hline
\end{tabular}
}
\caption{ \it Signal yield ($S$), background yield ($B$), and signal significance ($\mathcal{S}_{\rm sig}$) for different benchmark points under loose, medium, and hard selection criteria.}
\label{tab:significance_table}
\end{table}

The results in Table~\ref{tab:significance_table} exhibit a strong dependence on the mass splitting, $\Delta M = M_{E_1^\pm}-M_S$. Benchmark points with large mass splittings, such as BP1, BP2, BP4, BP5, BP7, and BP8, yield energetic electrons and substantial $\cancel{E}_T$, resulting in large $\mathcal{S}_{\rm sig}$ even after the application of stringent kinematic selections. For these benchmark points, the hard selection criteria provide the largest significances, reaching $\mathcal{S}_{\rm sig}=41.4$ for BP4 and $\mathcal{S}_{\rm sig}=33.8$ for BP5. In contrast, benchmark points with compressed spectra, namely BP3, BP6, and BP9, are characterized by softer leptons and reduced $\cancel{E}_T$. Consequently, the signal efficiency decreases rapidly with increasingly stringent cuts, leading to significantly lower sensitivities. Overall, the hard selection criteria are found to be optimal for benchmark points with large mass splittings, whereas looser selections are preferred in the compressed region of the parameter space. For each considered $(M_{E_1^\pm}, M_S)$ mass combination, the signal significance is computed for all three categories of kinematic cuts. The largest significance obtained is subsequently used in constructing the projected exclusion and discovery contours shown in the next subsection.

\subsection{Projected Exclusion Reach}

\begin{figure}[htbp]
\centering
\includegraphics[scale=0.4]{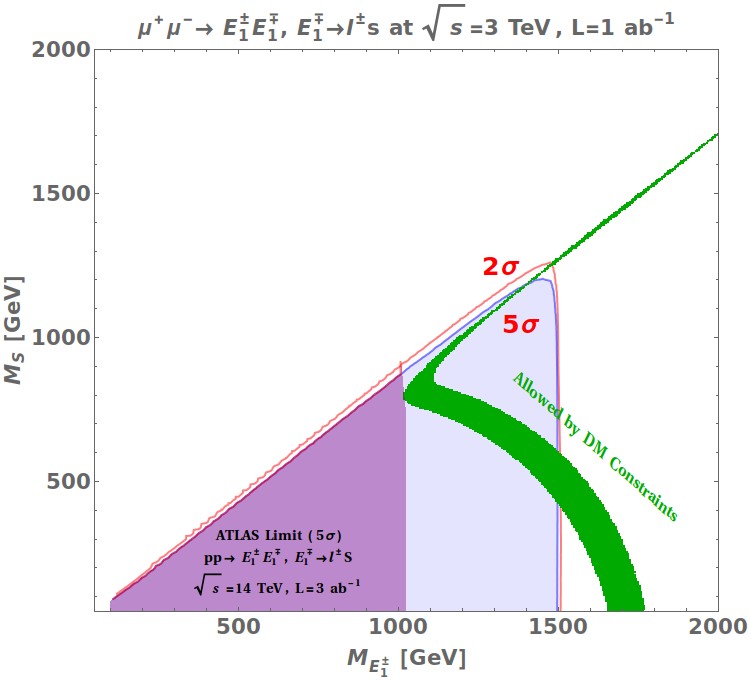}
\caption{ \it Projected $2\sigma$ exclusion (red line) and $5\sigma$ discovery reach (blue) in the $(M_{E_1^\pm},M_S)$ plane at a $\sqrt{s}=3$~TeV muon collider with an integrated luminosity of $1~\mathrm{ab}^{-1}$. The violet region indicates $5\sigma$ discovery reach at HL-LHC. The green region denotes the parameter space consistent with the dark matter constraints discussed in Sec.~\ref{sec:dm1}.}
\label{fig:exclusion_contour}
\end{figure}

Using the optimized selection strategy described in the previous subsection, we determine the projected exclusion and discovery sensitivities in the $(M_{E_1^\pm},M_S)$ parameter space. The resulting projected sensitivities at a $\sqrt{s}=3$~TeV muon collider with an integrated luminosity of $1~\mathrm{ab}^{-1}$ are shown in Fig.~\ref{fig:exclusion_contour}.

As evident from Fig.~\ref{fig:exclusion_contour}, a large region of the parameter space can be probed at a future muon collider. The projected $5\sigma$ discovery reach extends up to $M_{E_1^\pm}\simeq1.5$~TeV for favourable values of the dark matter mass, while the $2\sigma$ exclusion contour probes a slightly larger region. The sensitivity is strongest for scenarios with sizeable mass splittings, $M_{E_1^\pm}-M_S$, where the final-state electrons and missing transverse energy are sufficiently energetic to satisfy the selection requirements with high efficiency. The reach gradually decreases as $M_S$ approaches $M_{E_1^\pm}$. In this compressed region, the decay products become softer, leading to a reduction in both the electron transverse momentum and missing transverse energy distributions. Consequently, the signal efficiency decreases, resulting in weaker exclusion and discovery sensitivities near the kinematic boundary. Also shown in Fig.~\ref{fig:exclusion_contour} is the region consistent with the dark matter relic density, direct-detection bounds, and other phenomenological constraints discussed in Sec.~\ref{sec:dm1}. It can be seen that a substantial fraction of the dark-matter-allowed parameter space falls within the projected exclusion and discovery contours. In particular, charged fermion masses up to approximately $1.5$~TeV can be discovered over a broad range of dark matter masses. 

\begin{figure}[htbp]
\centering
\includegraphics[scale=0.4]{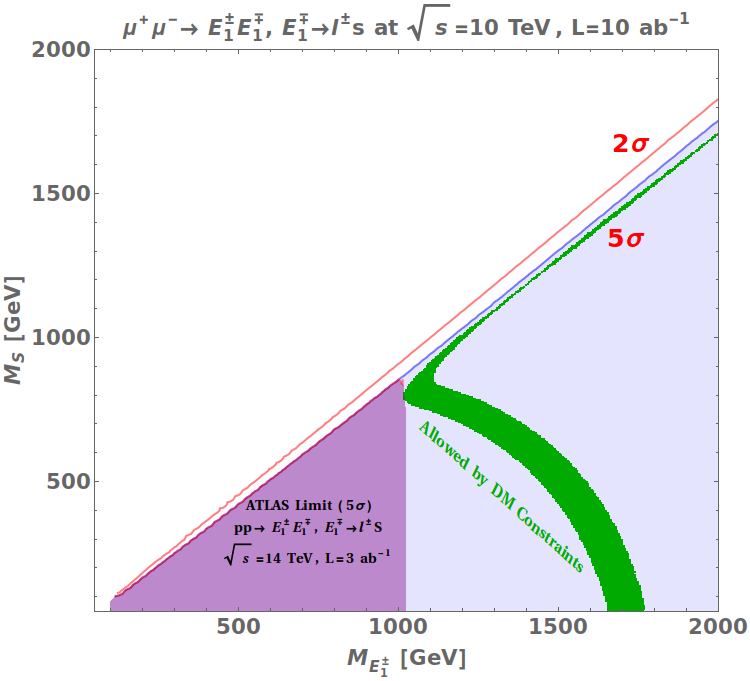}
\caption{ \it Projected $2\sigma$ exclusion (red line) and $5\sigma$ discovery reach (blue) in the $(M_{E_1^\pm},M_S)$ plane at a $\sqrt{s}=10$~TeV muon collider with an integrated luminosity of $10~\mathrm{ab}^{-1}$. The violet region indicates $5\sigma$ discovery reach at HL-LHC. The green region denotes the parameter space consistent with the dark matter constraints discussed in Sec.~\ref{sec:dm1}.}
\label{fig:exclusion_contour10}
\end{figure}

The corresponding results for a $\sqrt{s}=10$~TeV muon collider with an integrated luminosity of $10~\mathrm{ab}^{-1}$ are shown in Fig.~\ref{fig:exclusion_contour10}. The increase in centre-of-mass energy and integrated luminosity substantially enhances the sensitivity to heavier charged fermions. As shown in Fig.~\ref{fig:exclusion_contour10}, the projected $5\sigma$ discovery reach extends up to approximately $M_{E_1^\pm}\simeq 2$~TeV, while the $2\sigma$ exclusion contour probes an even larger region of parameter space. Similar to the $\sqrt{s}=3$~TeV case, the sensitivity is strongest for sizeable mass splittings between the charged fermion and the scalar dark matter candidate, whereas the reach gradually weakens in the compressed region due to the softer visible decay products. A particularly noteworthy feature of Fig.~\ref{fig:exclusion_contour10} is the substantial overlap between the collider-accessible region and the parameter space favoured by dark matter constraints. The green band corresponding to regions consistent with the observed relic abundance and current direct-detection limits lies almost entirely within the projected $5\sigma$ discovery contour. Consequently, a high-energy muon collider operating at $\sqrt{s}=10$~TeV would be capable of probing nearly the entire phenomenologically viable parameter space shown in the figure.

For comparison, the violet shaded regions in Figs.~\ref{fig:exclusion_contour} and \ref{fig:exclusion_contour10} indicate the projected sensitivity obtained from the ATLAS search for electroweak production of sleptons and charginos in the $\ell^+\ell^-+\cancel{E}_T$ final state. Since the signal topology closely resembles the process considered in this work, the ATLAS results~\cite{ATLAS:2019lff} provide a useful benchmark for assessing the relative reach of hadron and muon colliders. It is evident that while the HL-LHC sensitivity is largely limited to charged fermion masses around $1$~TeV, the muon collider significantly extends the accessible mass range and, more importantly, covers a much larger fraction of the parameter space consistent with dark matter constraints. These results therefore demonstrate that future muon colliders offer a powerful and complementary probe of the charged fermion and scalar sectors of the model, with the potential to explore regions of parameter space that remain inaccessible to the HL-LHC.

\section{Conclusion}\label{sec:conc}
With the discovery of Higgs, the Standard Model (SM) of particle physics remains the profound bedrock in this field. However, despite of being successful it is unable to explain many cosmological and astrophysical evidences, such as existence of dark matter etc. In this article, we revisit a scenario consisting of a real singlet dark matter augmented by new singlet and doublet fermions. This additional inclusion of singlet-doublet fermions associated with new Yukawa interactions creates new $t$- and $u$- channels, which help maintain the current relic abundance of the DM of the universe, satisfying the constraints coming from direct detection bounds. We also perform comprehensive analysis of the model coming from theoretical requirements such as vacuum stability, unitarity etc; along with some experimental bounds from lepton flavour violation, anomalous magnetic moment etc. 

We find that the pure Higgs-portal scenario is strongly constrained by current dark matter direct-detection limits. In particular, the latest combined LUX-ZEPLIN limits exclude almost the entire parameter space up to dark matter masses of about 3 TeV that we have examined. Only the Higgs resonance region around the dark matter mass of $62$ GeV is viable. However, the inclusion of the new Yukawa interaction significantly enlarges the allowed parameter space. In this case, the observed relic abundance can be achieved for dark matter masses ranging from approximately $10$ GeV up to the unitarity mass bound. The correct relic density is obtained for Yukawa couplings in the range $Y_f \simeq 0.35$-$0.45$, while the Higgs-portal coupling $\kappa$ remains below the present direct-detection limits. In the allowed region, the relic density is primarily governed by $ t$- and $ u$-channel annihilation processes, with co-annihilation effects becoming increasingly important for multi-TeV dark matter masses.

Furthermore, we explored the collider implications of the model at future muon colliders through the pair production of the lightest charged fermion, $\mu^+\mu^- \rightarrow E_1^+E_1^- \rightarrow e^+e^-+\cancel{E}_T$. Performing a detailed detector-level analysis, including realistic SM backgrounds and optimized kinematic selections, we determined the projected exclusion and discovery reaches in the $(M_{E_1^\pm},M_S)$ parameter space. At a $\sqrt{s}=3$ TeV muon collider with an integrated luminosity of $1~\mathrm{ab}^{-1}$, the $5\sigma$ discovery reach extends up to charged fermion masses of approximately $1.5$ TeV, while a $\sqrt{s}=10$ TeV muon collider with $10~\mathrm{ab}^{-1}$ can probe masses approaching $2$ TeV. Importantly, a large fraction of the parameter space compatible with relic-density and direct-detection constraints lies within the projected discovery reach, with the 10 TeV configuration capable of probing nearly the entire dark-matter-favoured region considered in this work. Compared to the projected HL-LHC sensitivity obtained from analogous dilepton plus missing-energy searches, future muon colliders provide substantially improved coverage of the viable parameter space and significantly extend the accessible mass range of the new charged fermions.

Overall, our study highlights the strong interplay between dark matter phenomenology and future collider searches. We demonstrate that the singlet scalar dark matter scenario supplemented by vector-like fermions remains a viable and well-motivated framework capable of simultaneously addressing the dark matter problem and providing experimentally testable signatures at next-generation collider facilities. The complementarity between direct-detection experiments and future muon colliders offers a powerful strategy for thoroughly exploring this scenario in the coming decades.


\section{Acknowledgements}
The work of S.~D.~is co-funded by the European Union’s Horizon Europe research and innovation program under the Marie Sk{\l}odowska-Curie COFUND Postdoctoral Programme grant agreement No. 101081355-SMASH and by the Republic of Slovenia and the European Union from the European Regional Development Fund. We appreciate AMU Maulana Azad Library for providing the licensed plagiarism and AI-spell checker Turnitin~\cite{turnitin} and Grammarly~\cite{grammarly2026}.



\bibliographystyle{utphys}
\bibliography{tevportalnew}

\newpage

\end{document}